\documentclass[floatfix,twocolumn,showpacs,preprintnumbers,amsmath,amssymb,pre,superscriptaddress]{revtex4-1}
\usepackage{color}
\usepackage[usenames,dvipsnames,svgnames,table]{xcolor}
\usepackage[colorlinks=true,linkcolor=blue,urlcolor=blue,citecolor=blue]{hyperref}

\usepackage{graphicx}
\usepackage{dcolumn}
\usepackage{bm}
\usepackage{subfigure}
\usepackage{amssymb}
\usepackage{multirow}
\usepackage{amsmath}
\usepackage{braket}
\graphicspath{{plots/}}

\begin{document}

\title{The Static Local Field Correction of the Warm Dense Electron Gas:\\ An \textit{ab Initio} Path Integral Monte Carlo Study and Machine Learning Representation}

\author{T.~Dornheim}
\email{t.dornheim@hzdr.de}

\affiliation{Center for Advanced Systems Understanding (CASUS), G\"orlitz, Germany}

\affiliation{Institut f\"ur Theoretische Physik und Astrophysik, Christian-Albrechts-Universit\"at zu Kiel,
 Leibnizstra{\ss}e 15, D-24098 Kiel, Germany}

\author{J.~Vorberger}

\affiliation{Helmholtz-Zentrum Dresden-Rossendorf, Bautzner Landstra{\ss}e 400, D-01328 Dresden, Germany}

\author{S.~Groth}

\affiliation{Institut f\"ur Theoretische Physik und Astrophysik, Christian-Albrechts-Universit\"at zu Kiel,
 Leibnizstra{\ss}e 15, D-24098 Kiel, Germany}

\author{N.~Hoffmann}

\affiliation{Helmholtz-Zentrum Dresden-Rossendorf, Bautzner Landstra{\ss}e 400, D-01328 Dresden, Germany}

\author{Zh.~A.~Moldabekov}

\affiliation{Institute for Experimental and Theoretical Physics, Al-Farabi Kazakh National University,  Al-Farabi Str.~71,  
  050040 Almaty, Kazakhstan}
  
\affiliation{Institut f\"ur Theoretische Physik und Astrophysik, Christian-Albrechts-Universit\"at zu Kiel,
 Leibnizstra{\ss}e 15, D-24098 Kiel, Germany}

\author{M.~Bonitz}

\affiliation{Institut f\"ur Theoretische Physik und Astrophysik, Christian-Albrechts-Universit\"at zu Kiel,
 Leibnizstra{\ss}e 15, D-24098 Kiel, Germany}

\begin{abstract}
The study of matter at extreme densities and temperatures as they occur in astrophysical objects and state-of-the art experiments with high-intensity lasers is of high current interest for many applications. While no overarching theory for this regime exists, accurate data for the density response of correlated electrons to an external perturbation are of paramount importance. In this context, the key quantity is given by the \textit{local field correction} (LFC), which provides a wave-vector resolved description of exchange-correlation effects.
In this work, we present extensive new path integral Monte Carlo (PIMC) results for the static LFC of the uniform electron gas, which are subsequently used to train a fully connected deep neural network. This allows us to present a continuous representation of the LFC with respect to wave-vector, density, and temperature covering the entire warm dense matter regime. 
Both the PIMC data and neural-net results are available online. Moreover, we expect the presented combination of \textit{ab initio} calculations with machine-learning methods to be a promising strategy for many applications.
\end{abstract}

\maketitle

\section{Introduction}

The uniform electron gas (UEG)~\cite{loos,quantum_theory}, often denoted as \textit{jellium}, is one of the most fundamental model systems in quantum many-body theory. Having been originally introduced as a simple model for conduction electrons in metals~\cite{mahan}, it offers a plethora of interesting physical phenomena such as Wigner crystallization~\cite{wigner,gs2,drummond_wigner,ichimaru_wigner,trail_wigner}, collective excitations~\cite{pines,pines1,pines2}, the possibility of a charge-/spin-density wave~\cite{iyetomi_cdw,holas_rahman,schweng,overhauser}, and an incipient excitonic mode at low density~\cite{takada1,takada2,dornheim_dynamic,higuchi}. Moreover, the accurate parametrization of its zero-temperature properties (e.g., Refs.~\cite{vwn,perdew,perdew_wang,gori-giorgi1,gori-giorgi2,new_param}) based on ground-state quantum Monte Carlo (QMC) simulations~\cite{gs2,gs1,moroni2,spink,ortiz1,ortiz2} has been pivotal for many applications, most notably as a basis for the striking success of density functional theory (DFT) regarding the description of real materials~\cite{dft_review,burke_perspective}.

Of particular importance is the response of the UEG to an external perturbation, which, within linear response theory, is fully captured by the density response function~\cite{after_burke} 
\begin{eqnarray}\label{eq:define_G}
\chi(q,\omega) = \frac{\chi_0(q,\omega)}{1 - \frac{4\pi}{q^2} \left[1-G(q,\omega)\right] \chi_0(q,\omega) }\ ,
\end{eqnarray}
 with $q$ and $\omega$ being the respective wave number and frequency. Here $\chi_0(q,\omega)$ denotes the response function of the ideal (i.e., noninteracting) system, and the local field correction (LFC) $G(q,\omega)$ contains all \textit{exchange-correlation} effects~\cite{kugler1}. Consequently, finding an accurate theory for the LFC has been a long-standing problem for decades~\cite{kugler1,hubbard}. Such information is crucial for many applications, including the interpretation of experiments~\cite{dominik,plagemann,fortmann1,fortmann2,neumayer}, the construction of effective, electronically screened potentials~\cite{ceperley_potential,zhandos1,zhandos2}, the development of advanced functionals for DFT~\cite{burke_ac,lu_ac,patrick_ac,goerling_ac}, the incorporation of electronic correlations into quantum hydrodynamics~\cite{diaw1,diaw2,zhanods_hydro}, and the calculation of other material properties like electrical and thermal conductivities~\cite{Desjarlais:2017,Veysman:2016}, stopping power~\cite{Cayzac:2017,Fu:2017}, and energy transfer rates~\cite{jan_relax}.

Consequently, a set of accurate data for the density response function and local field correction at zero temperature has been obtained on the basis of ground state-QMC simulations~\cite{moroni2,moroni,bowen2} in the static limit [i.e., $G(q,0)=G(q)$], which was subsequently used as input for the widespread analytical parametrization of $G(q)$ by Corradini \textit{et al.}~\cite{cdop}.

However, over the last years there has emerged a high interest in states of matter at extreme densities ($r_s=\overline{a}/a_\textnormal{B}\sim1$, with $\overline{a}$ and $a_\textnormal{B}$ being the average particle distance and first Bohr radius) and temperatures ($\theta=k_\textnormal{B}T/E_\textnormal{F}\sim1$, with $E_\textnormal{F}$ being the Fermi energy~\cite{quantum_theory}) as it occurs in astrophysical objects like giant planet interiors~\cite{militzer1,knudson,militzer3,manuel} and brown dwarfs~\cite{saumon1,saumon2,becker}.
This so-called \textit{warm dense matter} (WDM) can be realized and studied experimentally, e.g., via shock compression~\cite{koenig,fortov_review}, see Ref.~\cite{falk_wdm} for a topical review article.
Moreover, hot electrons have become important for many other communities, such as solid-state physics with laser excitations~\cite{ernstorfer2,ernstorfer}, high-pressure physics~\cite{diamond_anvil}, and hot-electron chemistry~\cite{hot_electron1,hot_electron2}. Further, WDM is predicted to materialize on the pathway towards inertial confinement fusion~\cite{hu_ICF,fortov_review,kritcher} and is important for the study of radiation damage cascades in the walls of fission and fusion reactors~\cite{damage}.
Therefore, WDM has emerged as one of the most active frontiers in plasma physics and materials science.

From a theoretical perspective, the intriguingly intricate interplay of 1) Coulomb coupling, 2) thermal excitations, and 3) quantum degeneracy effects prevents the application of perturbation expansions and renders the modelling of WDM a most formidable challenge~\cite{wdm_book}. In particular, the extension of ground state methods like DFT to these conditions~\cite{mermin_dft,rajagopal_dft} requires accurate and readily available knowledge of the fundamental properties of electrons in the warm dense regime. This need has sparked a surge of new developments in the field of electronic QMC simulations at finite temperature~\cite{brown_rpimc,schoof_cpimc,HF_nodes,dornheim,malone1,dornheim2,groth,dornheim3,dornheim_prl,malone2,dubois,dornheim_pre,dornheim_cpp,dornheim_neu,dornheim_pop}, which recently culminated in the highly accurate parametrization of the exchange-correlation free energy of the UEG covering the entire WDM regime~\cite{groth_prl,ksdt}, see Ref.~\cite{review} for a topical review article.

While being an important step in the right direction, a consistent and fundamental theory of WDM requires to go beyond the local density approximation~\cite{lda1,lda2}, with the electronic density response [see Eq.~(\ref{eq:define_G})] being of key importance. In this work, we aim to partly meet this demand by presenting a full description of the static local field correction $G(q;r_s,\theta)$ of the warm dense electron gas including all exchange-correlation effects. It is important to note that the results presented in this work correspond to the \textit{exact static limit} of $G(q,\omega)$ and have not been obtained on the basis of a \textit{static approximation} like many dielectric theories~\cite{stls,stls2,tanaka_new}. Moreover, we stress that $G(q)$ already constitutes a good approximation for $G(q,\omega)$ in many cases, as was recently demonstrated for the dynamic structure factor $S(q,\omega)$ in Refs.~\cite{dornheim_dynamic,dynamic_folgepaper}.

\begin{figure}
\includegraphics[width=0.41\textwidth]{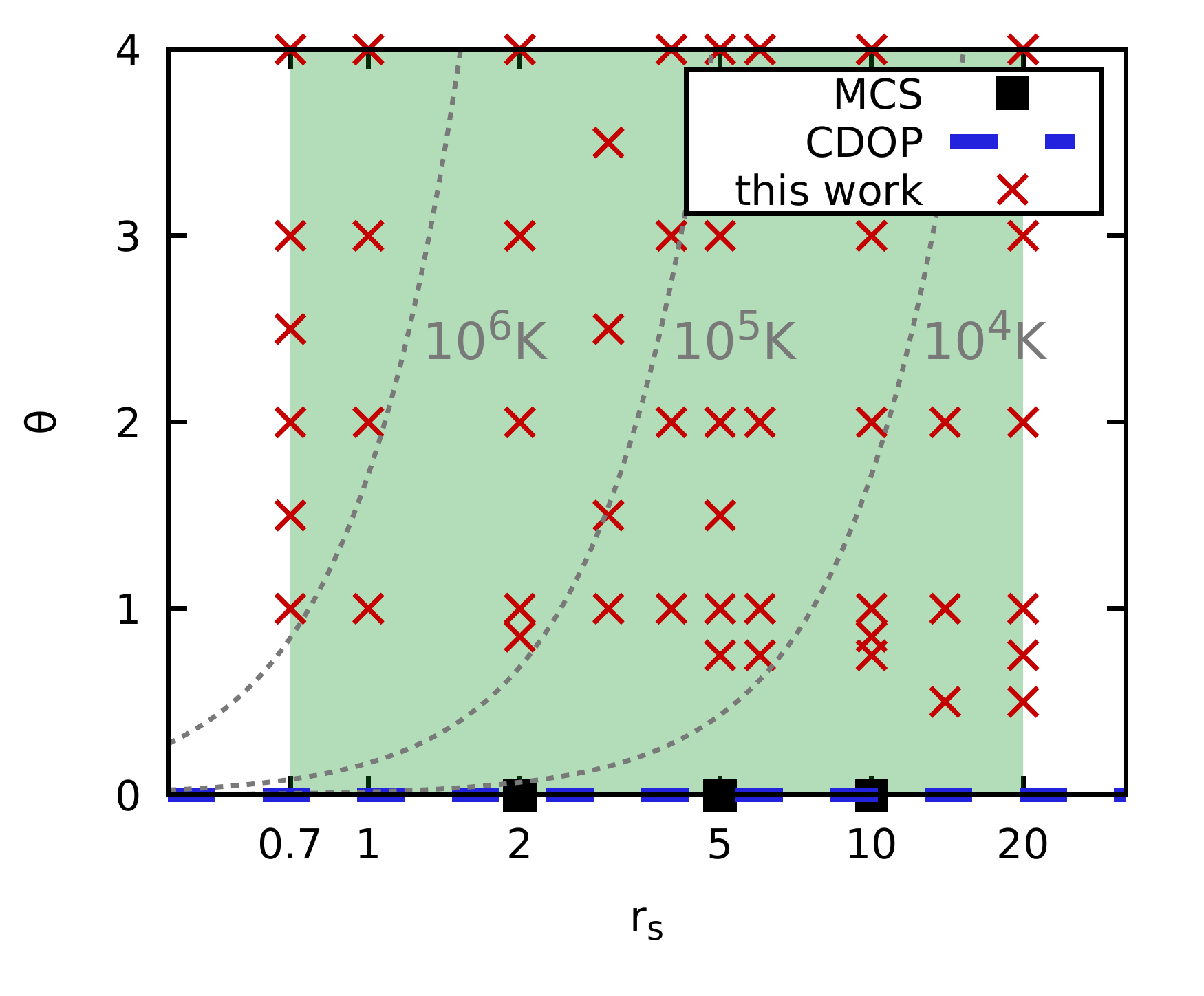}
\caption{\label{fig:PARAMETER_OVERVIEW}
Parameter overview of the $r_s$-$\theta$ plane: The dashed blue line depicts the ground state parametrization by Corradini \textit{et al.}~\cite{cdop} (CDOP) and the black squares the diffusion Monte Carlo points from Ref.~\cite{moroni2} (MCS) on which it is based. The red crosses correspond to our new PIMC data and the shaded green area depicts the parameter range for which our machine-learning representation (see Sec.~\ref{sec:representation}) has been constructed. The dotted grey curves indicate lines of constant temperature.
}
\end{figure}

In practice, we have carried out extensive path integral Monte Carlo (PIMC) simulations of the UEG at $N_\textnormal{p}\sim 50$ density-temperature combinations in the range of $0.7\leq r_s \leq 20$ and $0.5\leq\theta\leq4$, cf.~the red crosses in Fig.~\ref{fig:PARAMETER_OVERVIEW}.  
Here the main obstacle is given by the notorious fermion sign problem~\cite{dornheim_FSP,troyer}, which prevents PIMC simulations below half the Fermi temperature and limits the feasible system size to $N=14,\dots,66$. The latter issue can be controlled by using a previously introduced finite-size correction~\cite{groth_jcp}, such that any remaining difference to the thermodynamic limit ($N\to\infty$) vanishes within the given accuracy.
To overcome the restrictions regarding temperature, we combine our new PIMC data for $G(q)$ with the aforementioned ground-state parametrization (dashed blue line in Fig.~\ref{fig:PARAMETER_OVERVIEW}) based on the zero-temperature QMC data by Moroni \textit{et al.}~\cite{moroni2} (black squares).
These combined data are then used to \textit{train} a fully-connected deep neural net~\cite{NetNote}, which smoothly interpolates our data grid and provides accurate results for $G(q;r_s,\theta)$ at continuous densities and temperatures (see the green area in Fig.~\ref{fig:PARAMETER_OVERVIEW}) for up to five times the Fermi wave number, $0\leq q \leq 5q_\textnormal{F}$.

Both the new, machine-learning based representation of the static LFC and the extensive PIMC raw data have been made freely available online~\cite{LFC_git}, and can be used directly for the applications listed in the beginning. Moreover, the investigation both of the density response function $\chi(q)$ and $G(q)$ itself are interesting in their own right, and we find a pronounced dependence of the LFC on density and temperature, with a nontrivial behaviour around intermediate $q$-values and a negative large-$q$ tail for certain parameters. Furthermore, our results can be used to comprehensively gauge the accuracy of widely used approximations like dielectric theory~\cite{stls,stls2,tanaka_new,arora} and to benchmark and guide the developments of new methods~\cite{panholzer1}. Finally, we expect the presented combination of computationally expensive \textit{ab initio} calculations with modern machine-learning methods to be a promising strategy for many applications in many-body physics and beyond.

The paper is organized as follows: In Sec.~\ref{sec:workflow}, we introduce our \textit{ab initio} PIMC approach to compute the density response function $\chi(q)$, which allows to subsequently extract the desired local field correction. Moreover, we briefly discuss finite-size effects in our PIMC data, and demonstrate how they can be effectively removed. Sec.~\ref{sec:representation} is devoted to the machine-learning representation of $G(q;r_s,\theta)$, including discussions of the selected parameter range (\ref{sec:PR}) and the observed nontrivial behavior of $G$ with respect to density and temperature (\ref{sec:behave}). Finally, we briefly summarize our results and motivate their utility for future applications (\ref{sec:summary}).


\section{PIMC approach to the static density response\label{sec:workflow}}

We use a canonical adaption of the worm algorithm by Boninsegni \textit{et al.}~\cite{boninsegni1,boninsegni2} to carry out PIMC simulations of $N=2N_\uparrow=2N_\downarrow$ unpolarized electrons in a volume $V$ at a reduced temperature $\theta$. A detailed introduction of the PIMC method has been presented elsewhere~\cite{cep,review}. 
Further, we stress that we do not impose any nodal restrictions~\cite{fermion_nodes}, and our results are exact within the given Monte Carlo error bars.

Of particular importance in the context of the present work is
the imaginary-time density correlation function $F$ (see Ref.~\cite{dynamic_folgepaper} for details), which is defined as
\begin{eqnarray}\label{eq:F}
F(q,\tau) = \frac{1}{N} \braket{\rho(q,\tau)\rho(-q,0)}\ ,
\end{eqnarray}
with $\rho(q,\tau)$ being the density operator. \begin{figure}\centering
\hspace*{-0.093cm}\includegraphics[width=0.45\textwidth]{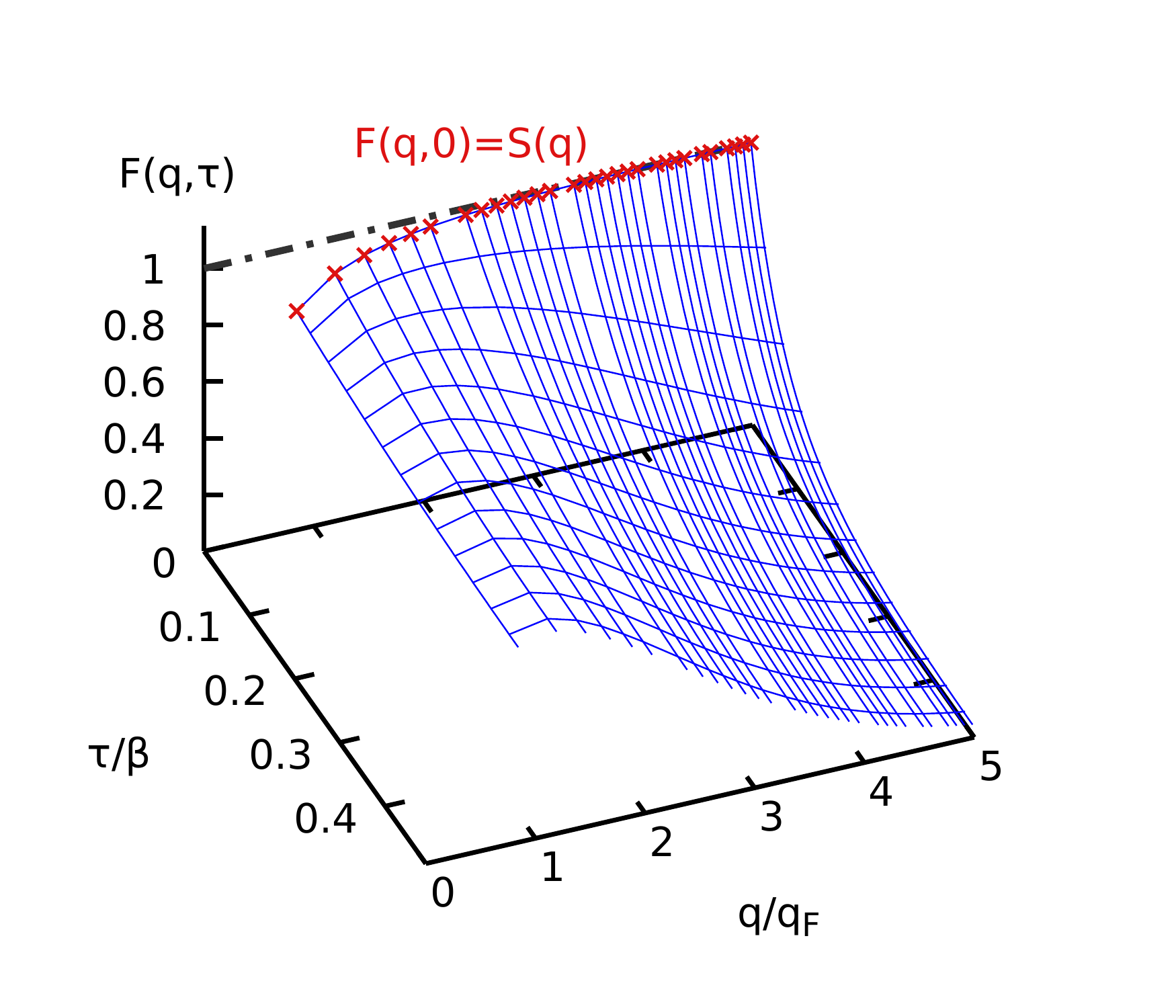} \\ \vspace*{-0.5cm}
\caption{\label{fig:ITCF}
PIMC results for the imaginary-time density-density correlation function $F(q,\tau)$ for $N=14$ unpolarized electrons at $r_s=1$ and $\theta=2$. For $\tau=0$, it is equal to the static structure factor $S(q)$.
}
\end{figure} In Fig.~\ref{fig:ITCF}, we show PIMC results for Eq.~(\ref{eq:F}) for $N=14$ electrons at $r_s=1$ and $\theta=2$ over the $q$-$\tau$-plane. Since a direct physical interpretation of $F(q,\tau)$ is rather difficult, here we only mention that it approaches the static structure factor $S(q)$ in the limit $\tau=0$ and that it is symmetric with respect to $\tau=\beta/2$ ($\beta=1/k_\textnormal{B}T$). The depicted $q$-grid is a direct consequence of the momentum quantization in a finite simulation cell, whereas the $\tau$-grid can, in principle, be made arbitrarily fine. We also note that $F(q,\tau)$ is connected to the dynamic structure factor $S(q,\omega)$ via a Laplace transform, which can be used as a starting point for the reconstruction of dynamic properties, see Refs.~\cite{dynamic_folgepaper,dornheim_dynamic}.

The main utility of $F$ in this work is the imaginary-time analogue of the fluctuation--dissipation theorem, which states that the static density response function can be computed as a simple one-dimensional integral along the $\tau$-axis~\cite{bowen,alex_prok},
\begin{eqnarray}\label{eq:static_chi}
\chi(q) = -n\int_0^\beta \textnormal{d}\tau\ F(q,\tau) \quad .
\end{eqnarray}
\begin{figure}
\includegraphics[width=0.41\textwidth]{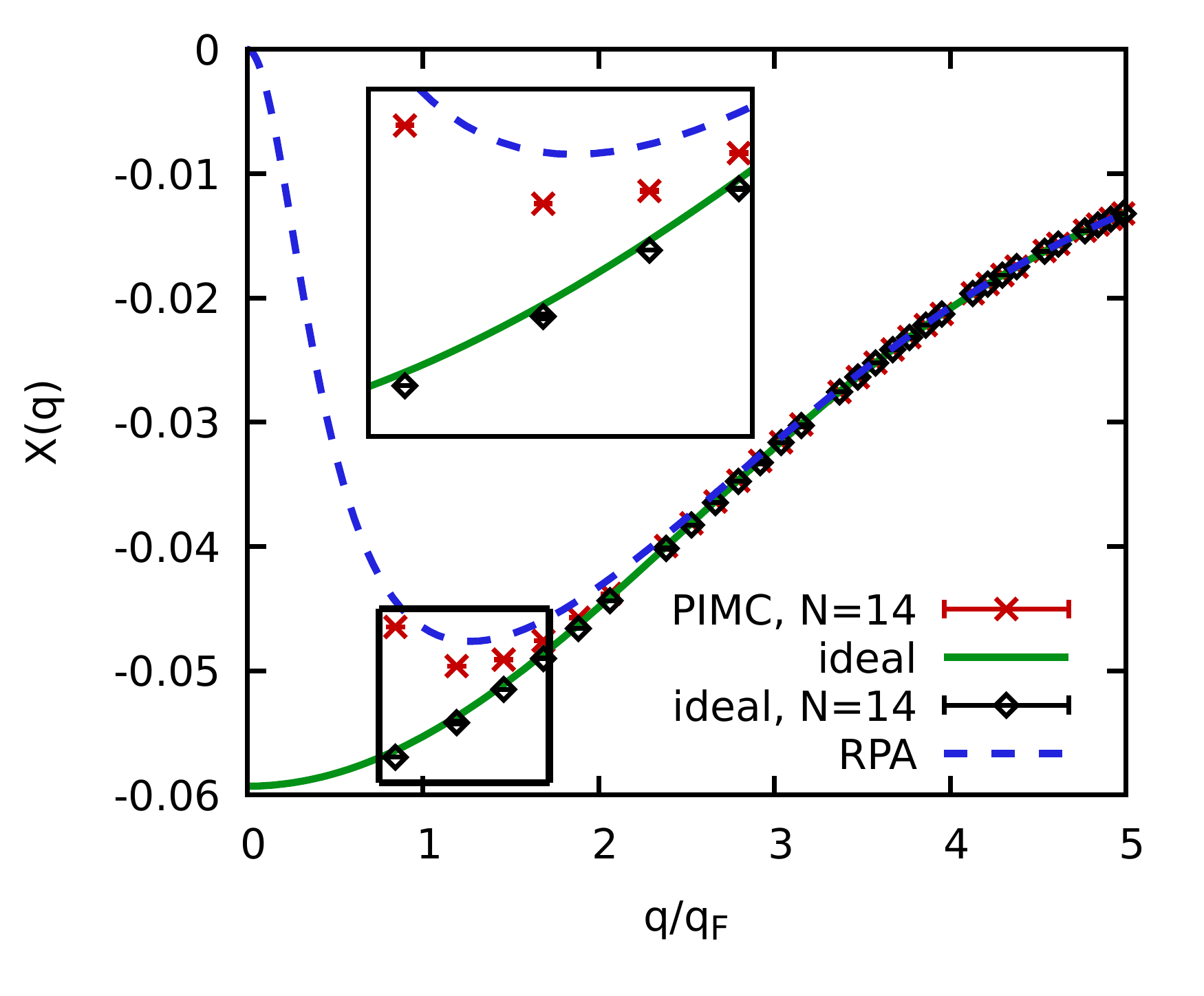}
\caption{\label{fig:chi_pic}
Wave-number dependence of the static density response function $\chi(q)$ for $r_s=1$ and $\theta=2$. The red crosses show PIMC results for $N=14$ computed as an integral over $F(q,\tau)$ [cf.~Eq.~(\ref{eq:F})] and the black diamonds the ideal response function for the same $N$ obtained from Configuration PIMC~\cite{groth_jcp}. The solid green and dashed blue lines correspond to $\chi_0(q)$ and $\chi_\textnormal{RPA}(q)$ in the TDL.
}
\end{figure} The results for Eq.~(\ref{eq:static_chi}) are shown in Fig.~\ref{fig:chi_pic} again for $r_s=1$ and $\theta=2$ as the red crosses for $N=14$. As a reference, we also include the widely used random phase approximation (RPA, dashed blue), which is obtained by setting $G(q,\omega)=0$ in Eq.~(\ref{eq:define_G}). This mean-field description of the static density response function is exact in the limits of $q\to0$ and $q\to\infty$, but significantly deviates in between. Furthermore, the solid green line corresponds to the ideal response function $\chi_0(q)$, which becomes exact for large $q$, but does not entail the screening effects that are manifest in the PIMC and RPA curves for small $q$~\cite{kugler2}.

The primary objective of this paper is the computation of the static LFC, which can be obtained from $\chi(q)$ by solving Eq.~(\ref{eq:define_G}) for $G(q)$, i.e.,
\begin{eqnarray}\label{eq:G_static}
G(q) = 1 - \frac{q^2}{4\pi}\left( 
\frac{1}{\chi_0(q)} - \frac{1}{\chi(q)}
\right)\ .
\end{eqnarray}
\begin{figure}
\includegraphics[width=0.41\textwidth]{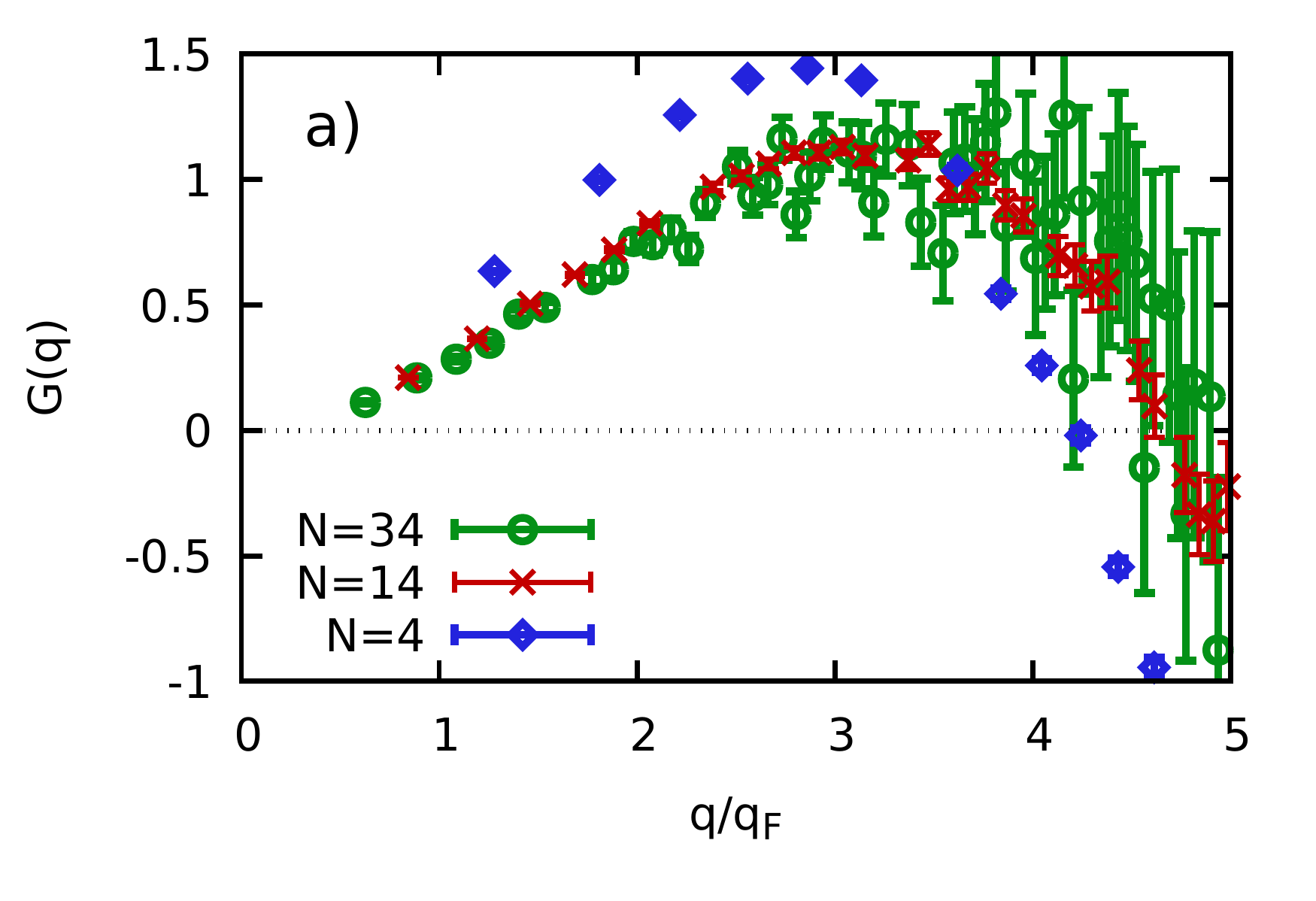}\\ \vspace*{-0.79cm}
\includegraphics[width=0.41\textwidth]{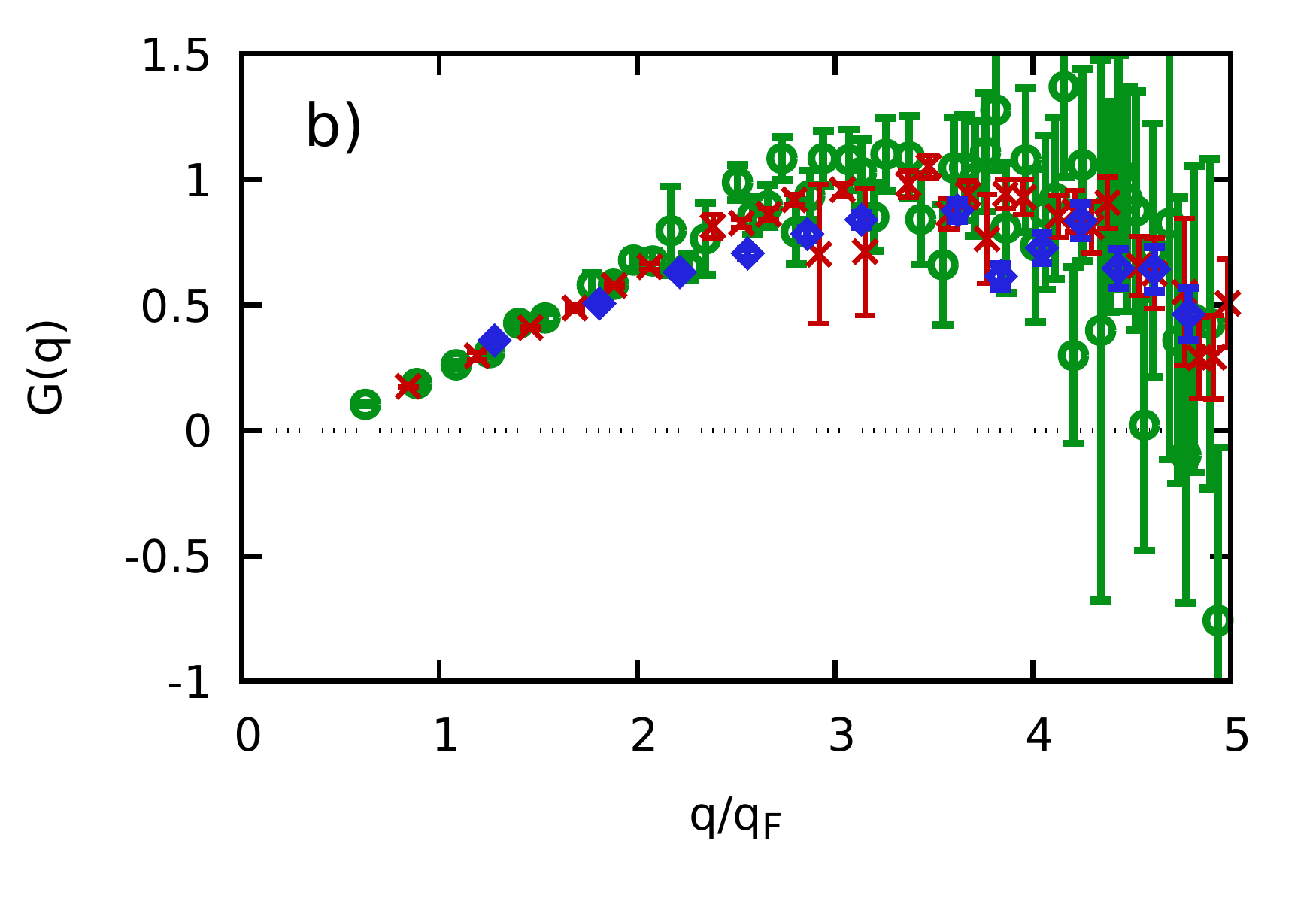}
\caption{\label{fig:G_pic}
Extracting the static local field correction: Panel a) shows PIMC data for $G(q)$ for $r_s=1$ and $\theta=2$ for $N=34$ (green circles), $N=14$ (red crosses), and $N=4$ (blue diamonds) unpolarized electrons computed from Eq.~(\ref{eq:G_static}) using $\chi(q)$. Panel b) shows the same data after the finite-size correction using $\chi_0^N(q)$.
}
\end{figure} The results for Eq.~(\ref{eq:G_static}) are shown in Fig.~\ref{fig:G_pic} a) for the same conditions as in Figs.~\ref{fig:ITCF} and \ref{fig:chi_pic} for three different particle numbers. 
First and foremost, we note that the error bars increase for large $q$. This is a direct consequence of the definition of $G(q)$ as a deviation measure between $\chi(q)$ and $\chi_0(q)$, cf.~Eq.~(\ref{eq:G_static}), which can be understood as follows: 
for small $q$, both functions are significantly different, and $G(q)$ can be accurately resolved. In contrast, they converge for large $q$ and the small remaining deviation is enhanced by the $q^2/4\pi$ pre-factor, which results in an increasing statistical uncertainty in the LFC. In practice, this means that we get accurate data for $G(q)$ when it has an important impact on the density response, whereas our estimation gets worse when $\chi(q)$ is essentially equal to the noninteracting response function in the first place.

Secondly, there are significant finite-size effects, which are particularly pronounced for $N=4$ (blue diamonds) as it is expected. 
The origin of this $N$-dependence in our PIMC data has been extensively discussed by Groth~\textit{et al.}~\cite{groth_jcp}, and is given by the system-size dependence of the ideal response function. This is illustrated by the black diamonds in Fig.~\ref{fig:chi_pic}, which depict configuration PIMC data for $\chi_0(q)$ for $N=14$ ideal fermions. While the deviations to the green curve are small, they do indeed become particularly significant when $\chi$ and $\chi_0$ converge. Since we are ultimately interested in a description of $G(q)$ in the thermodynamic limit, we finite-size correct our results by substituting the size-consistent ideal response function $\chi_0^N$ (which we computed using configuration PIMC, see Ref.~\cite{groth_jcp}) for $\chi_0$ in Eq.~(\ref{eq:G_static}), and the results are shown in Fig.~\ref{fig:G_pic} b).
Evidently, the system-size dependence has been drastically reduced even for as few as $N=4$ electrons, and the curves for $N=14$ and $N=34$ cannot be distinguished within the error bars. For completeness, we mention that the increased error bars at $N=34$ are a direct consequence of the fermion sign problem, see Ref.~\cite{dornheim_FSP} for an extensive discussion.

This procedure has been applied to all PIMC data shown in this work.

\section{Representation of the static local field correction\label{sec:representation}}

\begin{figure*}\vspace*{-0.9cm}
\hspace*{-0.99cm}\includegraphics[width=0.4\textwidth]{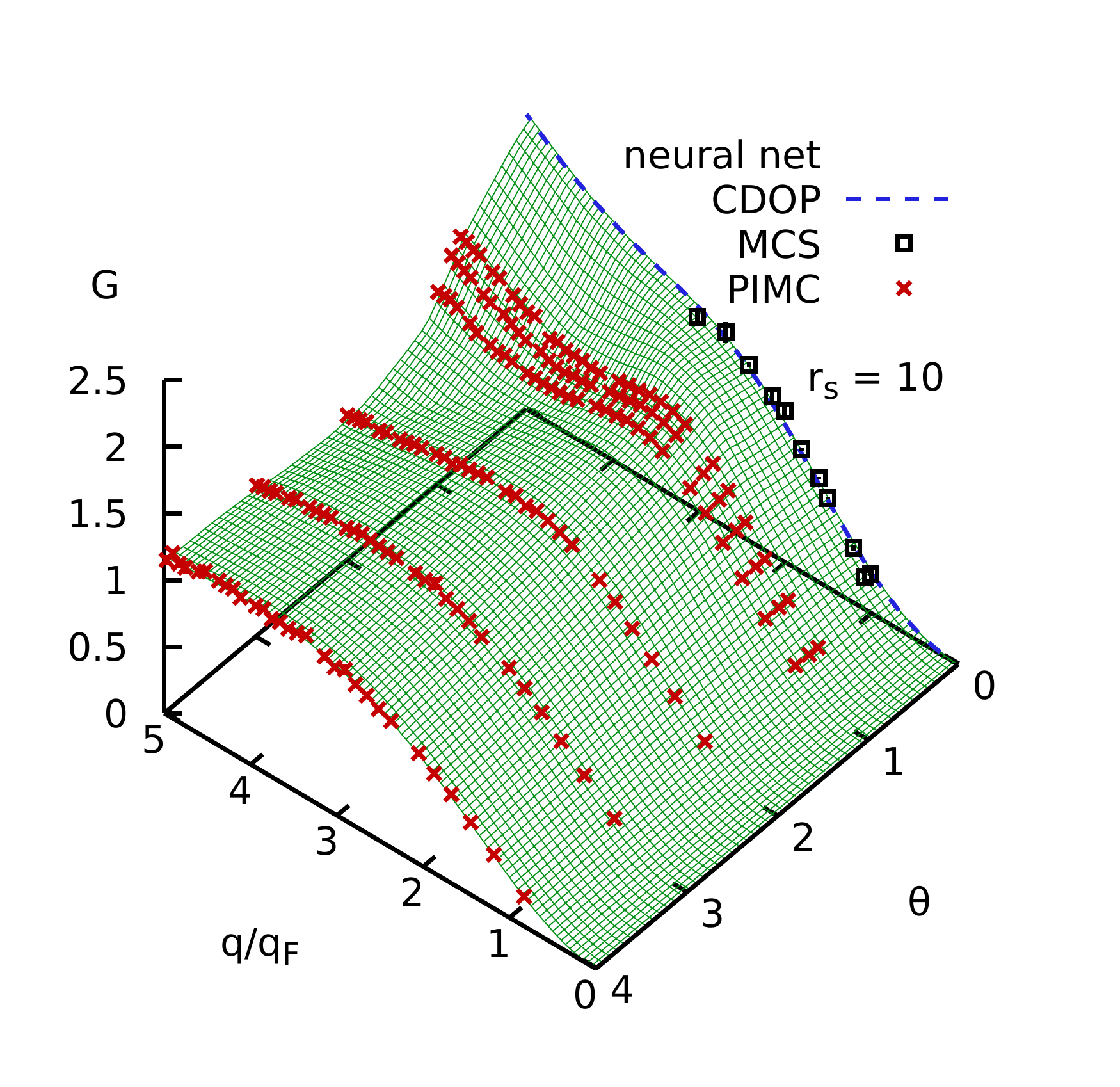}
\hspace*{-0.87cm}\includegraphics[width=0.4\textwidth]{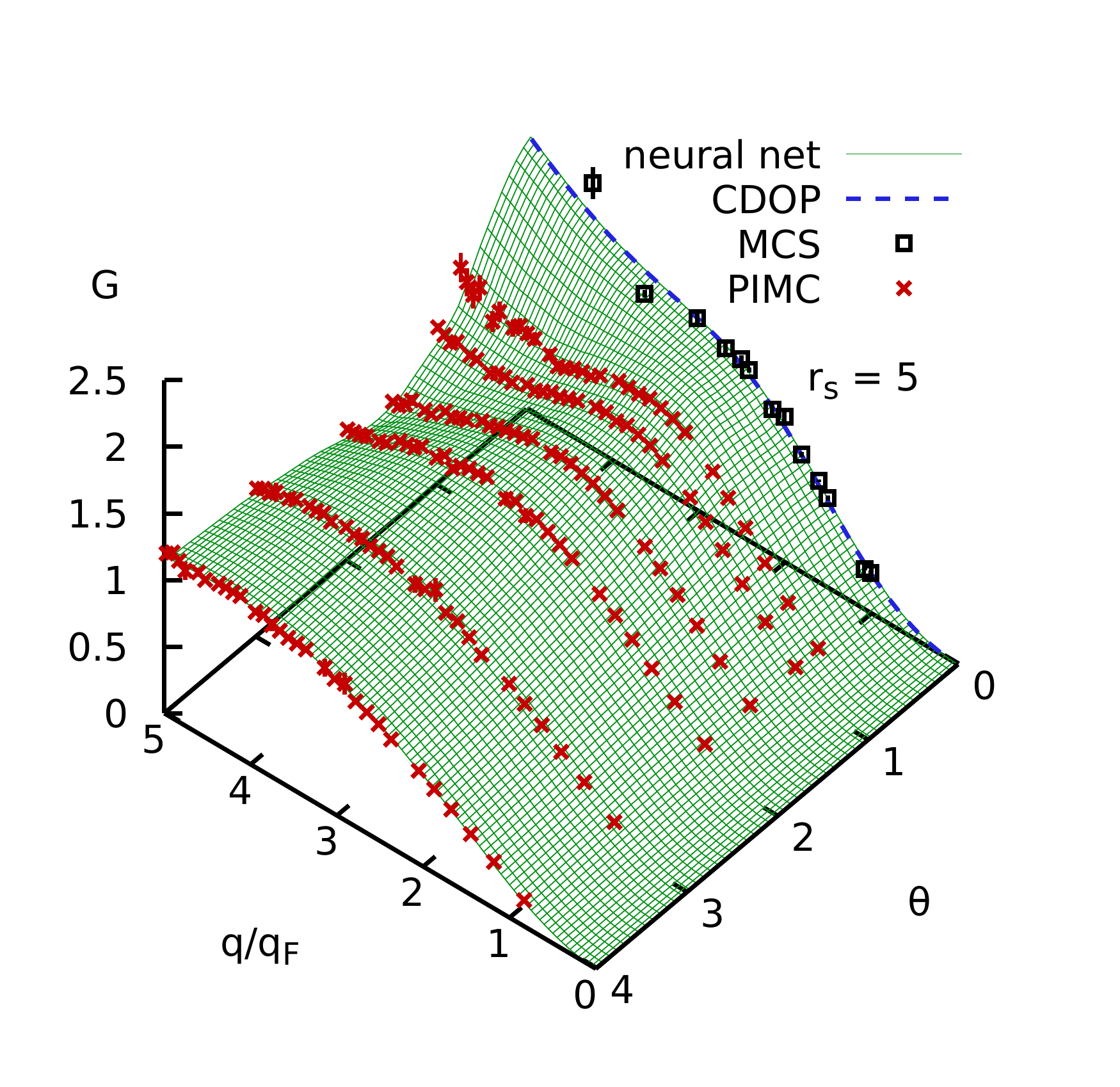}
\hspace*{-0.87cm}\includegraphics[width=0.4\textwidth]{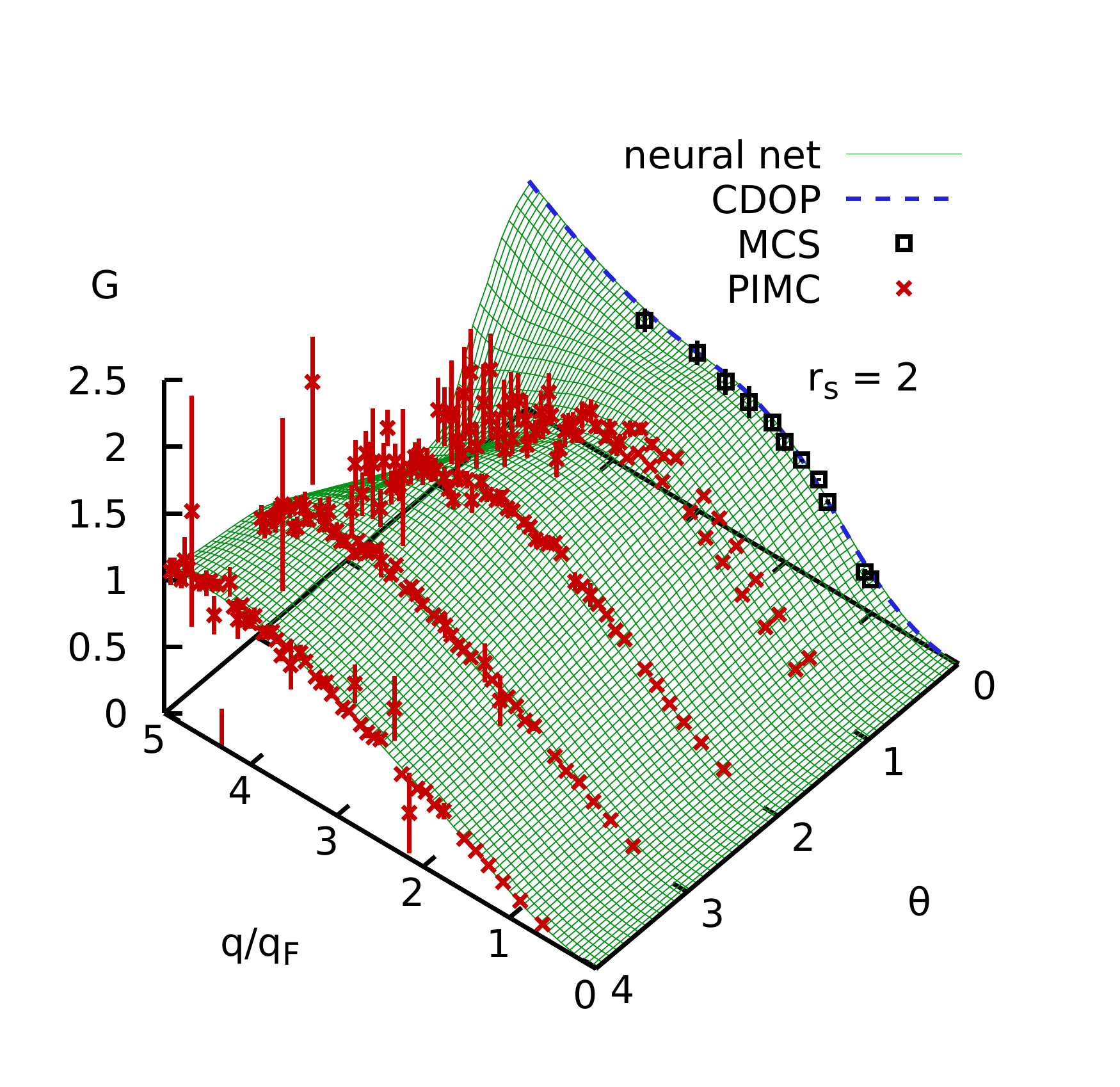} \vspace*{-0.59cm}
\caption{\label{fig:SPLOT_rs}
The static local field correction in the temperature-wave number plane for different values of the density parameter $r_s$: The dashed blue line depicts the ground state parametrization by Corradini \textit{et al.}~\cite{cdop} and the black squares the diffusion Monte Carlo points from Ref.~\cite{moroni2} on which it is based. The red crosses correspond to our new PIMC data computed from Eq.~(\ref{eq:static_chi}), and the green surface shows the machine-learning prediction.
}
\end{figure*}

The ultimate goal of this work is the construction of a continuous representation of the static local field correction with respect to wave number, density, and temperature, $G(q;r_s,\theta)$.
To this end, we have repeated the workflow introduced in Sec.~\ref{sec:workflow} for $N_\textnormal{p}\sim50$ different density-temperature combinations, and, to exclude the possibility of finite-size effects, often for different particle numbers $N$.

\subsection{Parameter range\label{sec:PR}}

Let us first discuss the available range of wave numbers $q$. Due to the aforementioned momentum quantization in a finite simulation cell, PIMC data are only available down to a minimum value of $q_\textnormal{min}=2\pi/L$, with $L=V^{1/3}$ being the box length. This, however, does not constitute a problem since the exact $q\to0$ limit is known from the compressibility sum-rule, which states that~\cite{stls2}
\begin{eqnarray}\label{eq:CSR}
\lim_{q\to0}G(q) = - \frac{q^2}{4\pi} \frac{\partial}{\partial n^2} \left( n f_\textnormal{xc} \right)\ ,
\end{eqnarray}
with $n$ being the density. In practice, Eq.~(\ref{eq:CSR}) is evaluated using the accurate parametrization of the exchange-correlation free energy $f_\textnormal{xc}$ from Ref.~\cite{groth_prl}.

The large-$q$ behaviour, on the other hand, is more difficult. As explained in Sec.~\ref{sec:workflow}, the relative uncertainty of our PIMC data increases with $q$, and $q=5q_\textnormal{F}$ does constitute a practical limit in many cases. Furthermore, the exact asymptotic expression for $q\to\infty$ introduced by Holas~\cite{holas_limit} only holds for $\theta=0$ and must not simply be extended to finite temperature~\cite{TailNote}. Therefore, we presently restrict ourselves to the range $0\leq q \leq 5q_\textnormal{F}$, which is fully sufficient for all practical applications.

The second relevant parameter range is the reduced temperature $\theta$. Since 1) the effect of $G(q)$ on $\chi(q)$ vanishes for large $\theta$ and 2) $G(q)$ eventually approaches the classical limit for any finite $q$, $\theta_\textnormal{max}=4$ constitutes a reasonable limit for WDM research. 
In addition, it is clear that a useful representation of $G(q;r_s,\theta)$ should extend down to the ground state. In this regard, we have chosen to incorporate the ground-state parametrization from Ref.~\cite{cdop}, and the validity range of our results is thus given by $0\leq \theta \leq 4$.

The last parameter range to be selected is the density parameter $r_s$. For small $r_s$ (i.e., high density), the system becomes weakly coupled. Hence, the influence of the local field correction vanishes and the RPA already provides an accurate description of the density response. In practice, we choose $r_s=0.7$ as the high-density limit of our representation, as it roughly coincides with core densities of giant planets~\cite{saumon1}.
Upon increasing $r_s$, the system becomes sparser and approaches a strongly coupled electron liquid. While this regime does not pose a challenge for our PIMC simulations, the selected range of $0.7 \leq r_s \leq 20$ is fully sufficient for WDM applications.

\subsection{Physical behavior and neural net representation\label{sec:behave}}

In the ground state, the static local field correction is relatively weakly dependent on $r_s$ and both the large- and small-$q$ limits are known analytically. Therefore, fitting an analytical representation to a QMC data set is a good strategy to obtain a smooth representation with only little bias~\cite{moroni2,cdop}. This situation, however, drastically changes at finite temperature. This can be seen in Fig.~\ref{fig:SPLOT_rs}, where we show $G$ in the $q$-$\theta$-plane for three different values of the density parameter $r_s$.

At zero-temperature, the LFC is of a parabolic form at small $q$ [cf.~Eq.~(\ref{eq:CSR})], exhibits a saddle point around $q\approx 3q_\textnormal{F}$, and finally reaches the asymptotic limit predicted by Holas~\cite{holas_limit}. In particular, it holds 
\begin{eqnarray}\label{eq:TAIL}
\lim_{q\to\infty}G(q;r_s,0) = B(r_s) + C(r_s) q^2 \ ,
\end{eqnarray}
where the constants $B$ and $C$ have been parametrized in Ref.~\cite{moroni2}. More specifically, the pre-factor $C$ is directly proportional to the change in the kinetic energy due to exchange-correlation effects~\cite{moroni2,holas_limit,farid}. Hence, $C$ is positive for all $r_s$, and $G(q)$ exhibits a positive tail at $\theta=0$.

At finite temperature, on the other hand, $G(q)$ behaves entirely different. At strong coupling (e.g., $r_s=10$, left panel) and high temperature, the UEG converges towards the classical one-component plasma and $G(q)$ becomes flat. 
This can be understood heuristically from Eq.~(\ref{eq:TAIL}) in the following way: for the classical system, the kinetic energy per particle attains the ideal value $K_\textnormal{cl}=3k_\textnormal{B}T/2$, and the pre-factor $C$ vanishes. Therefore, while Eq.~(\ref{eq:TAIL}) quantitatively holds only for $\theta=0$, it still predicts the correct qualitative behavior for $\theta>0$ in this case. 
Moreover, we note that our PIMC data for $G(q)$ at $r_s=10$ exhibit a small yet significant maximum at around $q\approx3q_\textnormal{F}$ followed by a minimum depending on temperature.

\begin{figure}\centering \vspace*{-0.6cm}
\includegraphics[width=0.4\textwidth]{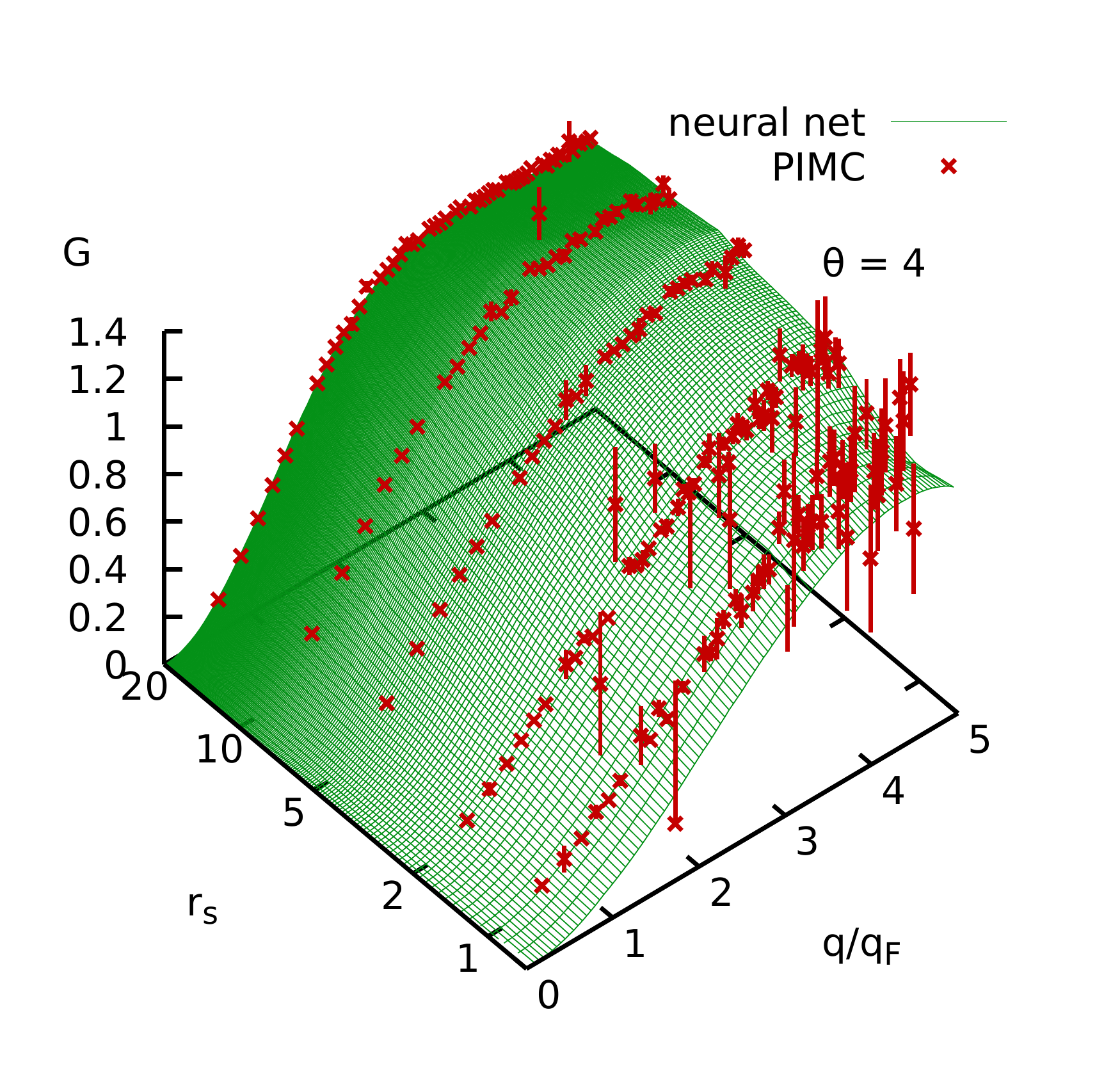}\\ \vspace*{-0.6cm}
\includegraphics[width=0.4\textwidth]{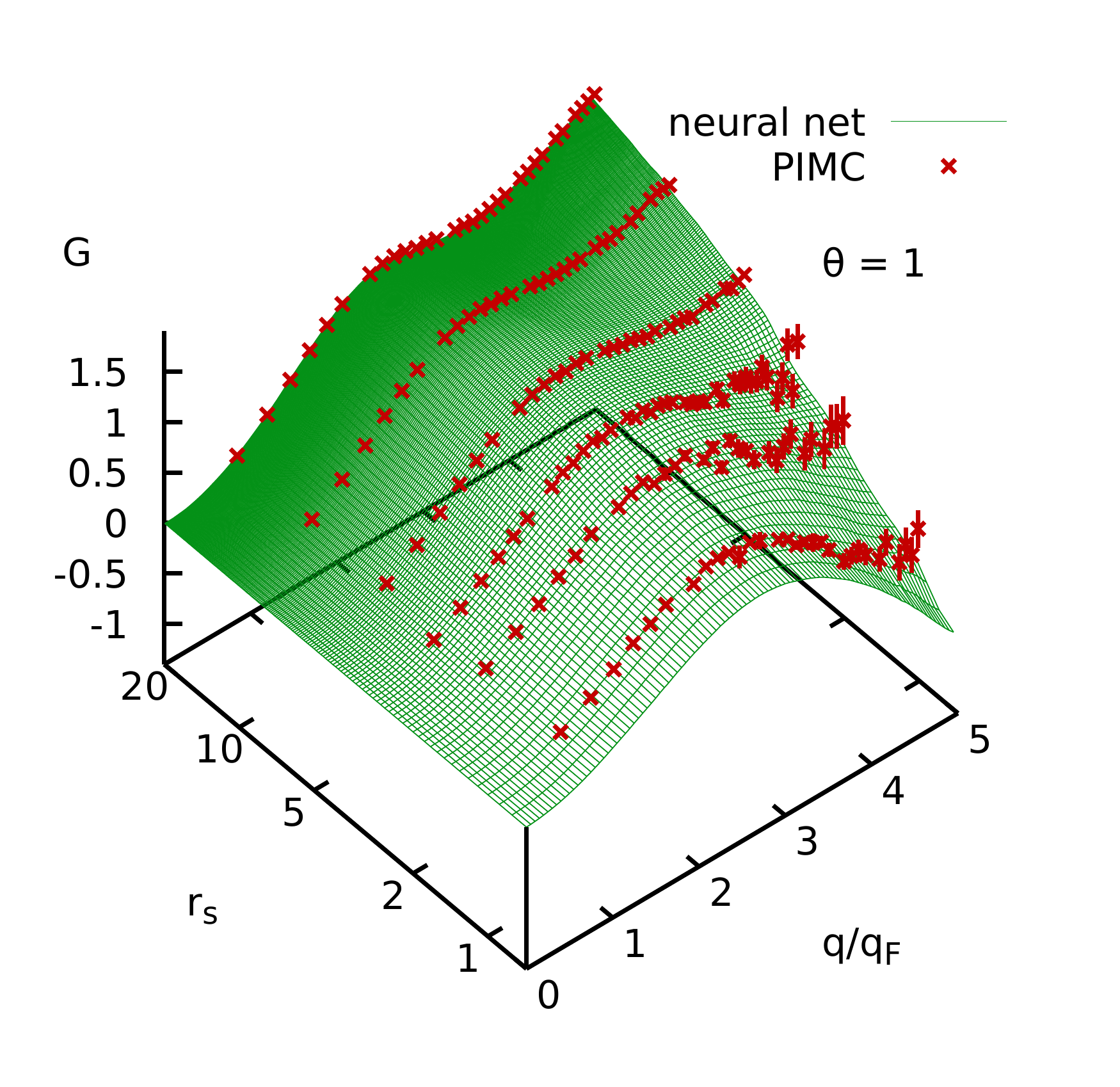} \vspace*{-0.36cm}
\caption{\label{fig:SPLOT_t1}
The static local field correction in the $r_s$-wave number plane for different values of the reduced temperature $\theta$: The red crosses correspond to our new PIMC data computed from Eq.~(\ref{eq:static_chi}), and the green surface shows the machine-learning prediction.
}
\end{figure}

At $r_s=5$ and $r_s=2$, which are located in the warm dense matter regime, the situation becomes even more complicated and, thus, interesting. While the large-$q$ behavior of $G(q)$ eventually always becomes flat for high temperature (for every finite value of $q$), and approaches Eq.~(\ref{eq:TAIL}) towards $\theta=0$, we observe \textit{negative tails} in the vicinity of the Fermi temperature. This can be seen even more clearly in Fig.~\ref{fig:SPLOT_t1}, where we show $G(q;r_s,\theta)$ in the $r_s$-$q$-plane. The top panel corresponds to $\theta=4$ and exhibits the expected classical, flat behavior at large $q$. The bottom panel depicts results for the Fermi temperature, and nicely illustrates the pronounced $r_s$-dependence of $G$ in the warm dense matter regime: at large $r_s$, $G(q)$ exhibits the by now familiar structure with a maximum, minimum, and a positive tail. With increasing density, the minimum becomes less distinct and eventually vanishes as the large-momentum tail becomes negative, and it clearly holds $G(q)<0$.

To understand this nontrivial finding, we might again consider Eq.~(\ref{eq:TAIL}), which predicts that the tail is parabolic with $K_\textnormal{xc}=K-K_0$ being the pre-factor. It has been known for some time~\cite{militzer_kinetic,kraeft_kinetic} that $K_\textnormal{xc}$ of the UEG does indeed become negative precisely in the WDM regime. While we again stress that the derivation of Eq.~(\ref{eq:TAIL}) is not valid at $\theta>0$ and that it does not quantitatively agree with our PIMC data, it is still likely that the change of the sign in $K_\textnormal{xc}$ is directly connected to the observed nontrivial behavior in $G(q;r_s,\theta)$.

In summary, we have found that the LFC does exhibit a distinct dependence on density and temperature, for which, despite being qualitatively understood, no analytical formula exists. Therefore, the utilization of an analytical fit formula would most likely introduce an artificial bias.
In contrast, modern machine-learning methods provide a flexible alternative in this situation and have emerged as a powerful tool for universal function approximation~\cite{NN1,NN2} that is not afflicted with this shortcoming.
In particular, we have used our extensive set of PIMC and CDOP data [cf.~Fig.~\ref{fig:PARAMETER_OVERVIEW}] to train a fully-connected deep neural network (see Appendix~\ref{sec:neural_net} for details), which can then be used to predict $G(q;r_s,\theta)$ at continuous values of all three parameters in the specified parameter range.

The results of this procedure are shown as the green surfaces in Figs.~\ref{fig:SPLOT_rs} and \ref{fig:SPLOT_t1}. 
First and foremost, we note the excellent agreement with the input data over the entire parameter range. More specifically, we find a mean deviation measure of 
\begin{eqnarray}
\chi = \frac{1}{N_\textnormal{data}} \sum_{k=1}^{N_\textnormal{data}}
|G_\textnormal{nn}^k-G^k_\textnormal{PIMC}| \approx 0.03 \quad .
\end{eqnarray}
In addition, the utilized weight regularization (cf.~Appendix~\ref{sec:neural_net}) brings about smooth curves with little generalisation errors, even in the most uncertain range of $0 < \theta < 0.5$.

\begin{figure}
\includegraphics[width=0.41\textwidth]{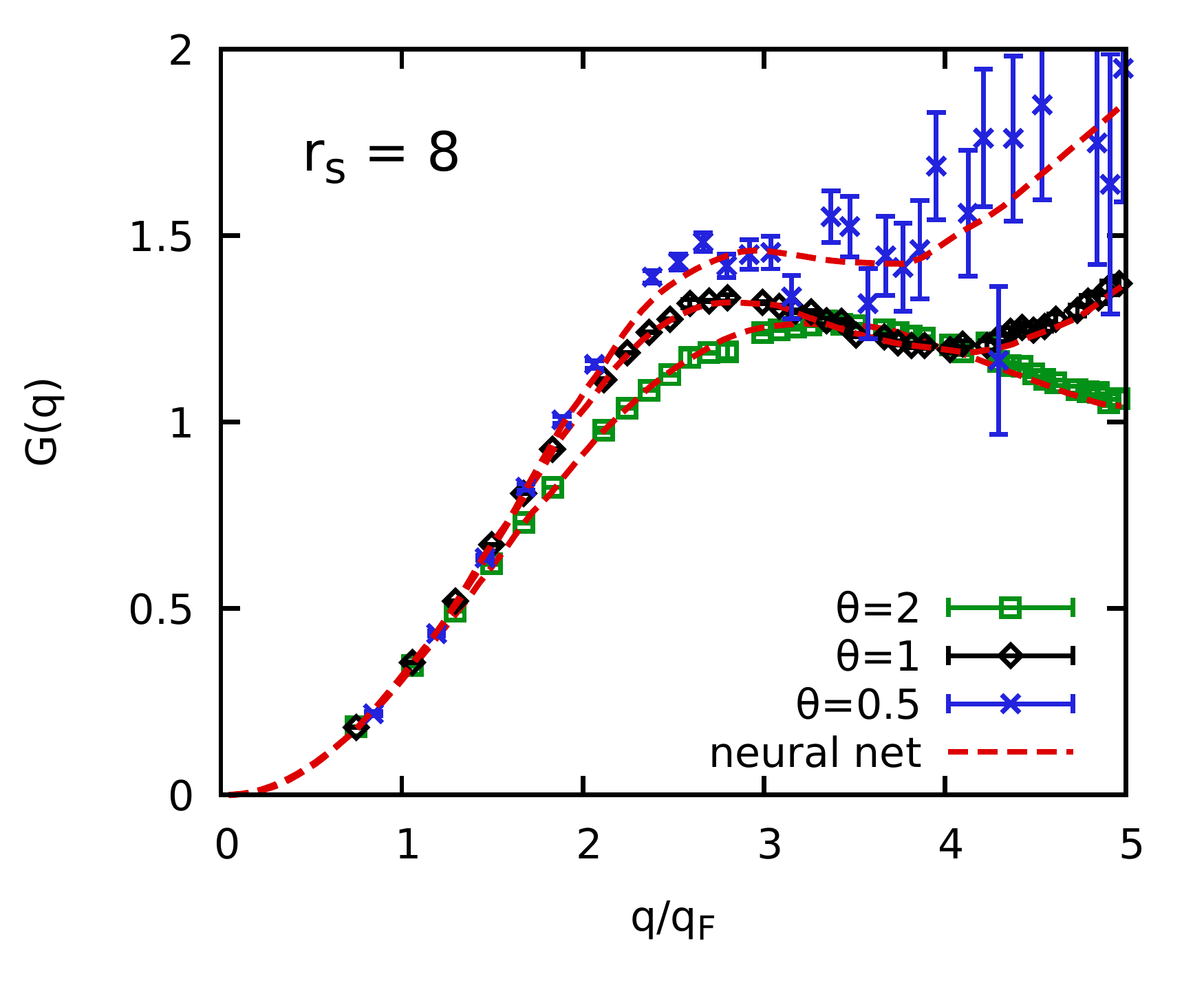}
\caption{\label{fig:prediction_pic}
Validation of the machine-learning representation: 
Shown is the wave-number dependence of $G(q)$  for three different temperatures at $r_s=8$. The different symbols correspond to finite-size corrected PIMC data that have not been included in the training of the neural net (cf.~Fig.~\ref{fig:PARAMETER_OVERVIEW}), and the red curves to the machine-learning prediction.
}
\end{figure} 
In Fig.~\ref{fig:prediction_pic}, we compare independent PIMC data for $r_s=8$ and three different temperatures that have not been included into the training of the neural net with the corresponding machine-learning prediction to validate the learned representation. The PIMC data for $\theta=2$ (green squares) and $\theta=1$ (black diamonds) are of high quality, and are in remarkable agreement with the neural net (red curves). At $\theta=0.5$, the simulations are computationally expensive due to the fermion sign problem (we find an average sign of $S\approx0.01$, see Ref.~\cite{dornheim_FSP} for an extensive discussion) and the statistical uncertainty is substantial for large $q$. Still, the neural net is capable to provide a prediction that is fully consistent with the benchmark data even in a regime where the training data are sparse.

\section{Summary and outlook\label{sec:summary}}

In this paper, we have presented extensive new PIMC results for the static local field correction $G(q)$ of the warm dense electron gas for $N_\textnormal{p}\sim50$ different density-temperature combinations and different particle numbers. These data---together with the ground-state parametrization from Ref.~\cite{cdop}---have subsequently been used as input to train a fully-connected deep neural network to learn the function $G(q;r_s,\theta)$. This has allowed us to obtain a fast, reliable, and continuous representation of the LFC covering the entire warm dense matter regime. More specifically, we avoid any artificial bias due to an insufficient analytical representation (only few exact properties of $G(q;r_s,\theta)$ are known analytically at finite $\theta$) and have carefully validated the machine-learning prediction against independent benchmark data.

In addition to the expected utility of our results for future applications (see below), the investigation of $G(q;r_s,\theta)$ is interesting in its own right, and we find a nontrivial and distinct dependence on density, temperature and wave number. In particular, $G(q)$ does exhibit a \textit{negative tail} at large $q$ in the vicinity of the Fermi temperature, which can most likely be attributed to a change in the sign of the exchange-correlation part of the kinetic energy.

We are confident that the presented full description of $G(q;r_s,\theta)$ will open up new avenues for many research applications. First and foremost, our representation can directly be used to study interesting properties of the UEG itself, like the possibility of a charge-density wave~\cite{iyetomi_cdw,holas_rahman,schweng} and instabilities~\cite{takada2}. Further, $G$ gives direct access to other important properties like the density-response function $\chi(q)$ [cf.~Eq.~(\ref{eq:define_G})] and the static dielectric function $\epsilon(q)$.
In addition, both our data and the neural net can be used to benchmark the accuracy of widespread approximations from dielectric theory like STLS~\cite{stls,stls2} and its recent improvement by Tanaka~\cite{tanaka_new}, and to guide the development of new approaches~\cite{panholzer1}.

While the present study of the LFC is restricted to the static limit, it has recently been shown that, even without taking the frequency dependence of $G(q,\omega)$ into account, many dynamical properties can be accurately estimated~\cite{dornheim_dynamic,dynamic_folgepaper}. For example, our representation of $G(q;r_s,\theta)$ can be directly used to compute the electronic dynamic structure factor $S(q,\omega)$ in a static approximation, which fully captures both the negative dispersion relation and the correlation induced broadening of the spectra due to exchange-correlation effects. Such readily available, yet accurate predictions of $S(q,\omega)$ are very valuable in many fields, most notably the interpretation of scattering experiments~\cite{dominik,siegfried_review}, which have emerged as a standard method of diagnostics in WDM experiments.

Moreover, we mention the utility of $G(q;r_s,\theta)$ as input for other simulations like quantum hydrodynamics~\cite{diaw1,diaw2,zhanods_hydro}, the construction of effective potentials~\cite{ceperley_potential,zhandos1,zhandos2}, and for the computation of other material properties like electrical and thermal conductivities~\cite{Desjarlais:2017,Veysman:2016}. Furthermore, our results can be used to construct advanced functionals for DFT based on the adiabatic-connection fluctuation-dissipation formalism~\cite{burke_ac,lu_ac,patrick_ac,goerling_ac}, or as the basis for the exchange-correlation kernel in time-dependent DFT~\cite{without_chihara}.

Lastly, we are convinced that the strategy to use machine-learning methods as a continuous representation of computationally expensive quantum Monte Carlo data is not limited to the present application, and can be further developed as a powerful paradigm in warm dense matter theory and beyond.

Both our extensive set of PIMC data and the machine-learning representation have been made freely available online~\cite{LFC_git}.


\vspace*{0.13cm}

 \section*{Acknowledgments}
 We gratefully acknowledge valuable feedback by M.~Bussmann. In addition, we thank T.~Sjostrom and S.~Tanaka for sending us the VS and HNC data shown in Fig.~\ref{fig:dielectric}.
This work has been partially supported by the Deutsche Forschungsgemeinschaft via project BO1366/13, 
by the Ministry of Education and Science of the Republic of Kazakhstan (MES RK) under grant no.~BR05236730,
and by the Center of Advanced Systems Understanding (CASUS) which is financed by Germany's Federal Ministry of Education and Research (BMBF) and by the Saxon Ministry for Science and Art (SMWK) with tax funds on the basis of the budget approved by the Saxon State Parliament.
We further acknowledge CPU time at the Norddeutscher Verbund f\"ur Hoch- und H\"ochleistungsrechnen (HLRN) via grant shp00015, at the clusters \textit{hypnos} and \textit{hemera} at Helmholtz-Zentrum Dresden-Rossendorf (HZDR), and at the computing centre of Kiel university.

\appendix

\renewcommand\thefigure{\thesection.\arabic{figure}}

\setcounter{figure}{0}  

\section{Neural net training and layout\label{sec:neural_net}}

To generate the training data for the neural network, we construct cubic basis splines $G^{r_s,\theta}_\textnormal{s}(q)$ connecting Eq.~(\ref{eq:CSR}) at small $q$ with our PIMC data for $q\geq 2\pi/L$. To prevent overfitting of the statistical noise, the smoothness of the splines $G_\textnormal{s}(q)$ is chosen such that
\begin{eqnarray}\label{eq:wchi}
\chi_w = \frac{1}{M} \sum_{i=1}^{M} \frac{\left|G^{r_s,\theta}(q_i)-G^{r_s,\theta}_\textnormal{s}(q_i)\right|}{\Delta G^{r_s,\theta}(q_i)} \geq 0.9\ ,
\end{eqnarray}with the $M$ input points containing both PIMC and CSR data.
\begin{figure}
\includegraphics[width=0.41\textwidth]{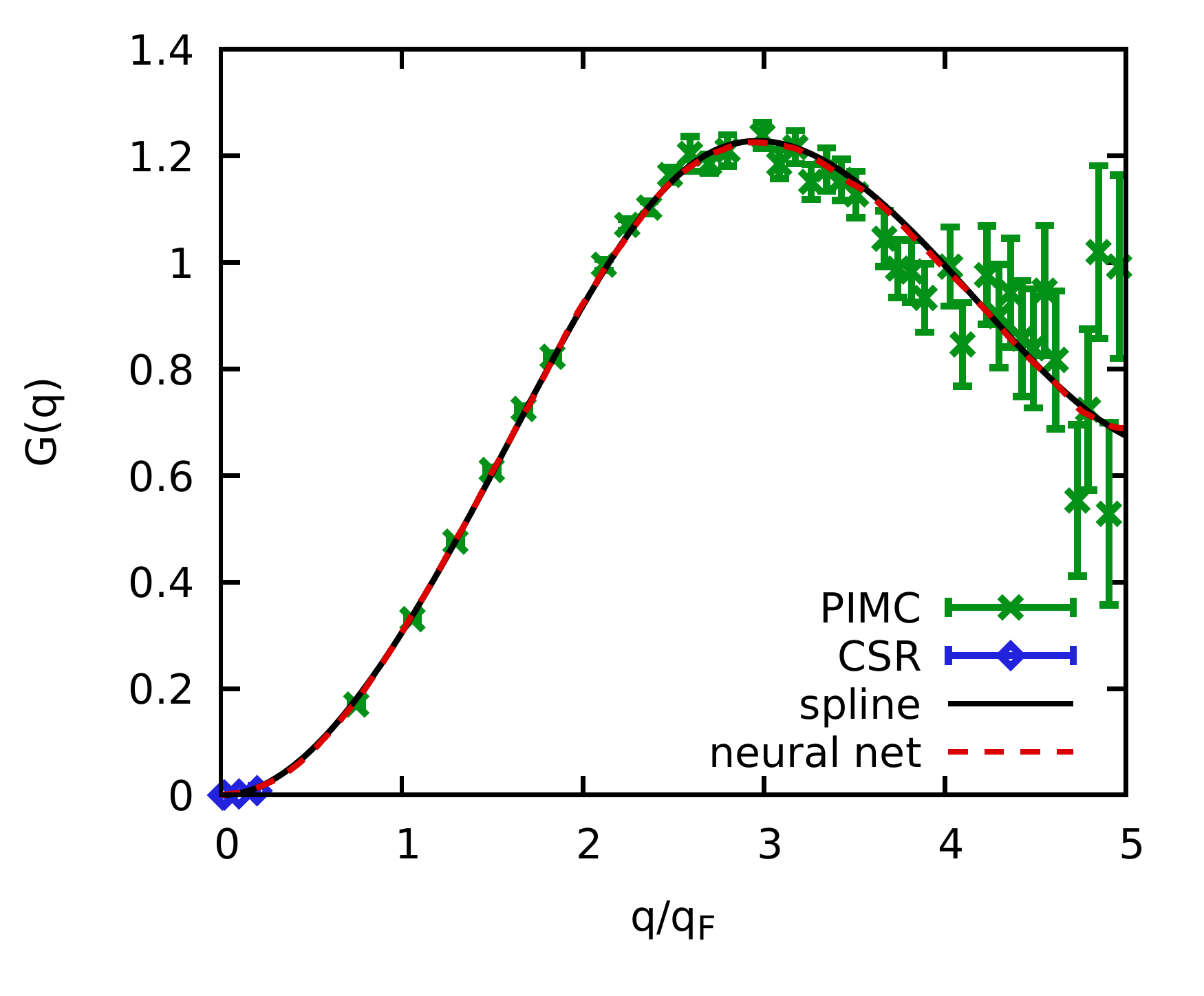}
\caption{\label{fig:spline}
Generation of the ML-input for $G(q;r_s,\theta)$ for $r_s=3$ and $\theta=1$. Shown are the finite-size corrected PIMC data (green crosses), points from the CSR [c.f.~Eq.~(\ref{eq:CSR}), blue diamonds] for small $q$, and a cubic basis spline (solid black). In addition, the dashed red line corresponds to the curve learned by the neural net.
}
\end{figure} This is illustrated in Fig.~\ref{fig:spline} for $r_s=3$ and $\theta=1$, with the spline (solid black) smoothly interpolating the somewhat noisy PIMC data (green crosses) and reproducing the CSR for small q (blue diamonds). This procedure has three main advantages: 1) evaluating the already smoothed spline to generate the training data makes the training of the net more stable, 2) it allows us to generate the training data on an equidistant $q$-grid, whereas the PIMC raw data become denser for large $q$, which could potentially bias the quality of the net, and 3) it stabilizes the net in the small intermediate $q$-range where neither PIMC nor CSR data are available (cf.~Fig.~\ref{fig:spline}).

In addition, we include CDOP~\cite{cdop} data at $\theta=0$ for $r_s=0.7,1,2,3,4,5,6,10,14,20$ into our training set. The total number of $G(q;r_s,\theta)$ samples is given by $N_\textnormal{train}\sim6.5\times10^{4}$.

\begin{figure}
\includegraphics[width=0.41\textwidth]{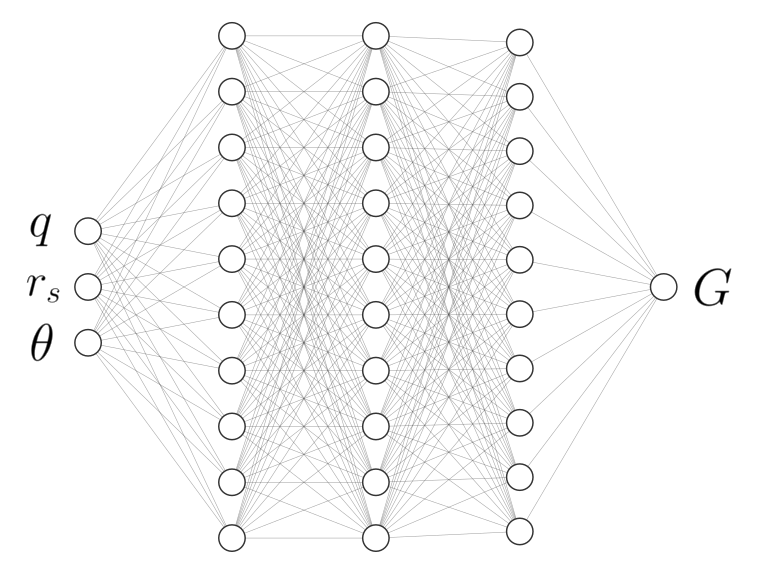}
\caption{\label{fig:layout}
Schematic illustration of the machine-learning representation of the static local field correction $G(q;r_s,\theta)$. The first layer is comprised of three neurons corresponding to the function arguments $q$, $r_s$, and $\theta$, and the final layer gives the prediction for $G(q;r_s,\theta)$. The net is fully connected and contains $N_\textnormal{l}=40$ hidden layers with $N_\textnormal{n}=64$ neurons each.
}
\end{figure} In Fig.~\ref{fig:layout}, we show a schematic illustration of the fully connected deep neural network representing the static LFC~\cite{NetNote}. The first layer corresponds to the three input variables $q$, $r_s$, and $\theta$, and the final layer provides the desired prediction of $G(q;r_s,\theta)$. Moreover, there are $N_\textnormal{l}=40$ hidden layers with $N_\textnormal{n}=64$ neurons each. This combination was found empirically and is sufficiently flexible to accommodate the complicated behavior of $G(q;r_s,\theta)$. In addition, we note that fully connected networks are particularly powerful regarding function approximation~\cite{NN1}.
We choose to optimize the neural network by ADAM~\cite{myADAM} and fixed its parameters to $\beta_1 = 0.9$, $\beta_2 = 0.999$, and learning rate $\lambda = 0.001$. Furthermore, we reduced the generalization error by layer-wise L2 regularization of weights (excluding bias) at $6\times10^{-7}$.

\begin{figure}
\includegraphics[width=0.41\textwidth]{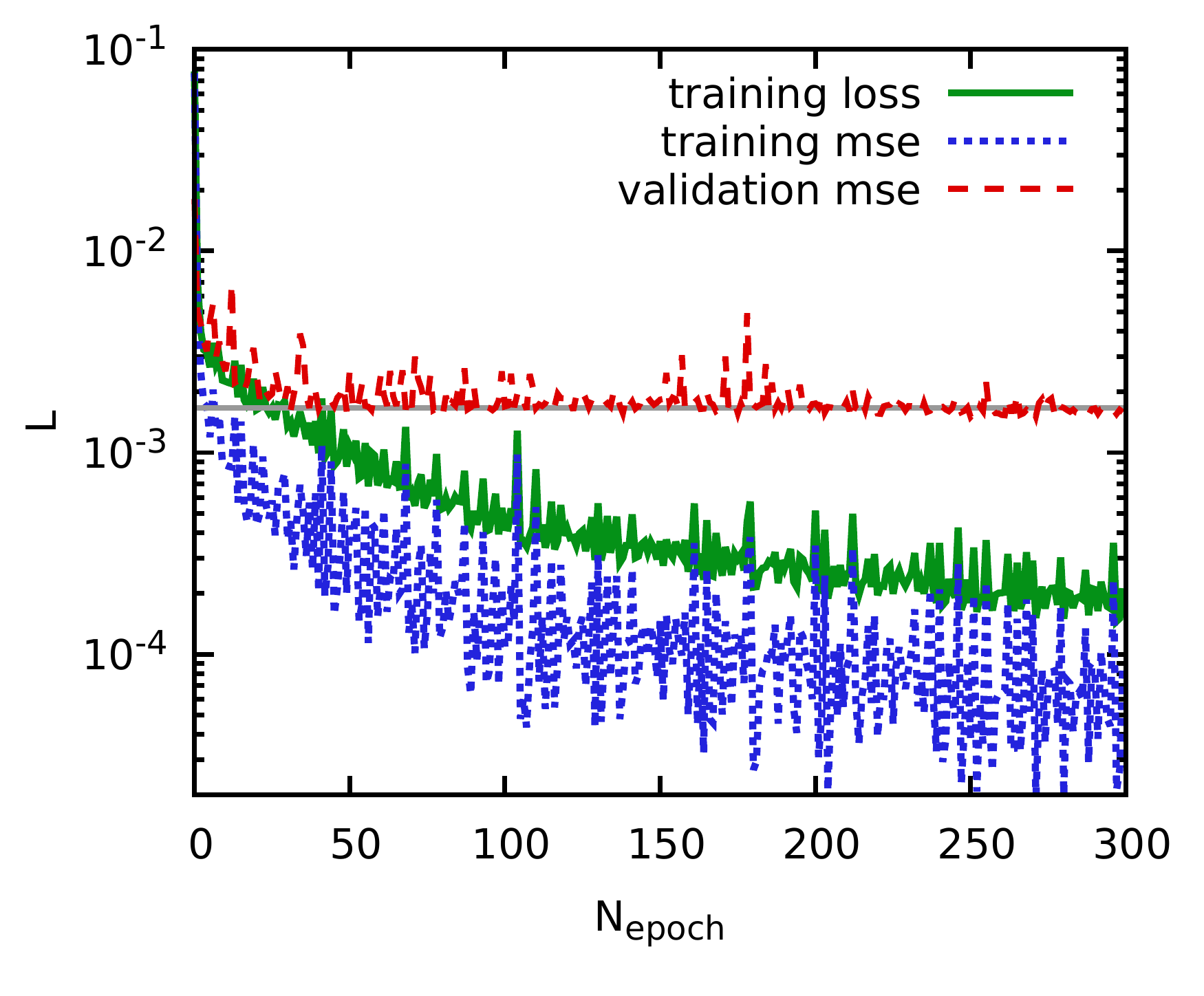}
\caption{\label{fig:loss}
Convergence of the training. The dotted blue and solid green lines show the mean-squared-error (mse) and the loss function (mse and weight regularization) as a function of the number of epochs. The dashed red line corresponds to the mse with respect to independent validation data ([$r_s=1.25$,$\theta=1.5$],[$r_s=8,\theta=0.5,0.75,1.0,2.0$], [$r_s=0.85,\theta=1.5$], and [$r_s=4,\theta=1.5$]), and the solid grey line to a constant fit for $200\leq N_\textnormal{epoch} \leq300$.
}
\end{figure}

Finally, the convergence behavior of the training is illustrated in Fig.~\ref{fig:loss}. The dotted blue line corresponds to the mean-squared-error (mse) with respect to the training data, and the solid green line to the actual loss function, which consists of both mse and the aforementioned weight-regularization. Therefore, the loss function is always larger than the pure mse. In addition, the red dashed line shows the mse computed with respect to independent validation data ($N_\textnormal{v}\sim8\times10^3$ samples for $G(q;r_s,\theta)$, see the caption of Fig.~\ref{fig:loss}) that were not included in the training. The solid grey line shows the average of the validation mse computed for $200 \leq N_\textnormal{epoch} \leq 300$, where the latter remains almost constant. This strongly implies the absence of a generalization error in our training procedure.

\begin{figure}
\includegraphics[width=0.41\textwidth]{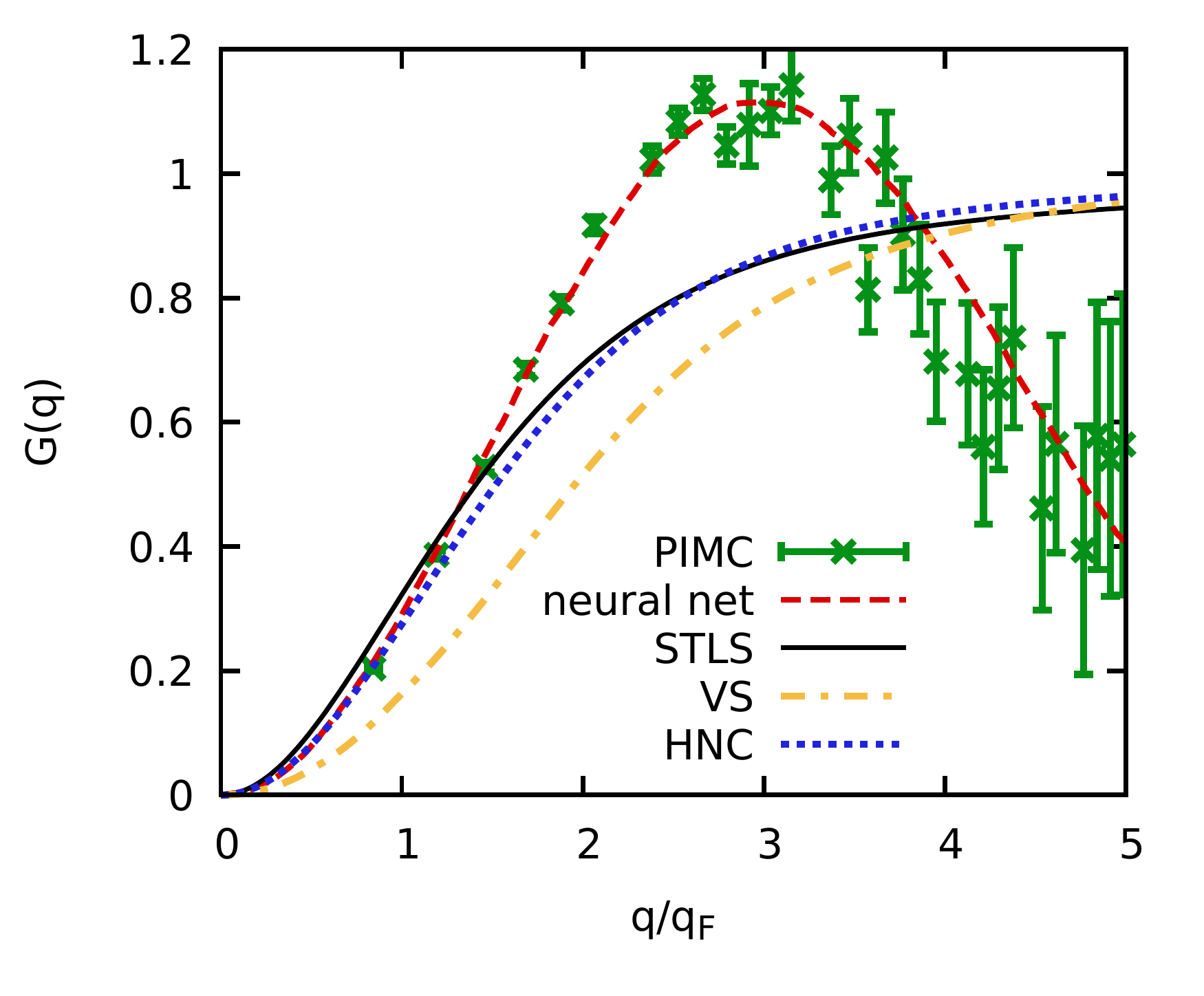}
\caption{\label{fig:dielectric}
Wave-number dependence of the static local field correction at $r_s=2$ and $\theta=1$. The green crosses and dashed red line correspond to our new finite-size corrected PIMC data and the machine-learning representation. In addition, we show several dielectric approximations, namely Singwi-Tosi-Land-Sj\"olander~\cite{stls} (STLS, solid black), Vashista-Singwi~\cite{stls2} (VS, dash-dotted yellow), and the recent hypernetted-chain based approach by Tanaka~\cite{tanaka_new} (HNC, dotted blue).
}
\end{figure} 

\section{Comparison to dielectric theory}

\setcounter{figure}{0}

In Fig.~\ref{fig:dielectric}, we compare our PIMC data (green crosses) and machine-learning representation (dashed red line) to various dielectric approximations for $r_s=2$ and $\theta=1$, i.e., at the center of the WDM regime. 
While the recent hypernetted-chain based approach by Tanaka~\cite{tanaka_new} constitutes a remarkable improvement over  STLS~\cite{stls} (solid black), and Vashista-Singwi~\cite{stls2} (dash-dotted yellow) at small $q$, no approximation is capable to qualitatively reproduce the full nontrivial behavior of $G(q)$ at these conditions. Therefore, the new \textit{ab initio} data presented in this work are indispensable to achieve a more fundamental understanding of the electronic density response.

A more detailed benchmark study of dielectric theory is beyond the scope of this work and will be presented elsewhere.



\begin{thebibliography}{10}
\bibitem{after_burke} We assume Hartree atomic units throughout this work.
\bibitem{NetNote} We have used the Keras~\cite{keras} frontend on the basis of TensorFlow~\cite{tensorflow}.
\bibitem{LFC_git} A link to  both the ML representation of the static LFC and the PIMC raw data for the density response will be provided upon publication.
\bibitem{TailNote} This is, among other things, due to the softening of the Fermi surface at finite temperature, and will be discussed in a separate publication.
\end{thebibliography}
\end{document}